\documentstyle[epsf,twocolumn,eqsecnum,aps,prb]{revtex}
\begin{document} 
\title{Surface magnetization and critical
behavior\\  of aperiodic Ising quantum chains}  
\author{Lo\"\i c Turban} 
\address{Laboratoire de Physique du Solide, 
Universit\'e de Nancy I, Bo\^\i te Postale 239,\\ 
F-54506  Vand\oe uvre l\`es Nancy Cedex, France} 
\author{Ferenc Igl\'oi}
\address{Research Institute for Solid State Physics, P.O. Box 49,\\ 
H-1525 Budapest 114, Hungary} 
\author{Bertrand Berche} 
\address{Laboratoire de Physique du Solide, 
Universit\'e de Nancy I, Bo\^\i te Postale 239,\\  
F-54506  Vand\oe uvre l\`es Nancy Cedex, France\\
{\sf (cond-mat/9312080)}\\
\vspace{4mm}
\parbox{14cm}{\rm\small\hspace{4mm}
We consider semi-infinite two-dimensional layered Ising models in the
extreme anisotropic limit with an aperiodic modulation of the
couplings. Using substitution rules to generate the aperiodic
sequences,  we derive functional equations for the surface
magnetization. These equations are solved by iteration and the critical exponent
$\beta_s$  can be determined exactly. The method is applied to three
specific aperiodic sequences, which represent different types of perturbation,
according to a relevance-irrelevance criterion. On the Thue-Morse lattice, for
which the modulation is an irrelevant perturbation, the surface magnetization
vanishes with a square root singularity, like in the homogeneous lattice. For 
the period-doubling sequence, the perturbation is marginal and $\beta_s$ is a
continuous function of the modulation amplitude. Finally, the
Rudin-Shapiro sequence, which corresponds to the relevant case,  displays an
anomalous surface critical behavior which is analyzed via scaling
considerations. Depending on the value of the modulation, the
surface magnetization either vanishes with an essential singularity or
remains finite at the bulk critical point, i.e., the surface phase transition is
of~first~order.
}}
\maketitle
\narrowtext
\section{Introduction}
\label{sec:intro}
Since the discovery of quasicrystals,\cite{shechtman84} one has witnessed
a growing interest  in understanding their structure and physical
properties (for recent reviews  see Refs.
\onlinecite{henley87,janssen88,janot89,guyot91,steinhardt91}). The
study of phase transitions on quasiperiodic lattices is now an active
field of research. Since quasiperiodic and, more generally, aperiodic
systems  can be considered as a state of  matter which interpolates
between the periodic (crystalline) and the random (disordered, glassy)
states, the corresponding critical behavior is very rich and
theoretically challenging.

The first results obtained in this field were mainly numerical. For
example, it was shown that different systems on the two-dimensional 
Penrose lattice (Ising model,\cite{godreche86,okabe88,sorensen91}
percolation,\cite{sakamoto89,zhang93} self-avoiding 
walks\cite{langie92}) are universal in the sense that the critical
exponents  are the same as in regular, two-dimensional lattices. The
same  conclusion was reached for three-dimensional quasiperiodic
structures,\cite{okabe90} although with a weaker numerical accuracy.

Exact results about critical properties can be obtained on two-dimensional,
layered Ising models, where the intralayer couplings $K_1$ are kept
constant, whereas the interlayer interactions $K_2(k)$ follow the 
aperiodic modulation of the underlying lattice. One usually works~in  
the~extreme~anisotropic limit,\cite{kogut79} with $K_2(k)\!\to\! 0$,
$K_1\!\to\!\infty$, keeping fixed the ratio $\lambda_k\!=\!
K_2(k)/K_1^*$ where  $K_1^*$ is the dual interaction corresponding to
$K_1$. In this limit  the system is described by a one-dimensional
quantum Ising model (QIM)  with the Hamiltonian 
\begin{equation}
{\cal H}=-{1\over 2}\sum_{k=1}^\infty\left[\sigma_k^z+\lambda_k
\sigma_k^x\sigma_{k+1}^x \right],\label{itf} 
\end{equation}
where the $\sigma$'s are Pauli spin $1/2$ operators.

The first exact results on the QIM were obtained for the bulk critical
properties on the Fibonacci and related lattices, where the 
specific heat is found to have a logarithmic singularity of the Onsager 
type.\cite{igloi88,doria88,benza89,henkel92,lin92} It was
Tracy\cite{tracy88} who first showed examples of aperiodic  layered Ising
models where the logarithmic singularity in the specific  heat is washed
out, like in the layered Ising model with random couplings 
(McCoy-Wu model).\cite{mccoywu}

A systematic study of the bulk critical behavior of aperiodic QIMs
has been performed recently by Luck.\cite{luck93a} Generalizing the
Harris  criterion,\cite{harris74} the relevance or irrelevance of the
aperiodicity is  shown to be connected to the size of the fluctuations
in the couplings  $\lambda_k$. For bounded fluctuations (as happens for
the Fibonacci  lattice) the specific heat of the system diverges
logarithmically as  in the homogeneous one, while, for unbounded
fluctuations, the critical  behavior is anomalous, e.g., the specific heat
displays an essential  singularity like in the McCoy-Wu
model.\cite{mccoywu} In the marginal case, when  the fluctuations  grow
on a logarithmic scale, non-universal critical  behavior is expected 
with critical exponents depending on the strength  of the aperiodicity.

The relevance-irrelevance criterion has been generalized for other 
systems\cite{igloi93} and higher dimensional
aperiodicities.\cite{luck93b}  For relevant modulations, the form of the
singular quantities  near the critical point has also been determined,
using scaling arguments.\cite{igloi93}

As far as the surface critical behavior of aperiodic systems is
concerned, only a few results are available. For the QIM in the ordered 
phase, the asymptotic limit of the spin-spin correlation function in
the surface gives the square of the surface magnetization $m_s$. This leads
to the expression\cite{schultz64}
\begin{equation}
m_s=<1\vert\sigma_1^x\vert 0>
\label{mat-el}
\end{equation}
where $\vert 0>$ is the ground state of $\cal H$ and $\vert 1>$ is the
first excited state, belonging to the odd sector of the Hamiltonian, which
is degenerate with the ground state in the ordered phase of the infinite
system. The matrix element in Eq.~(\ref{mat-el}) can be rewritten using a
Jordan-Wigner transformation of the spin operator, followed by a canonical
transformation to Fermi operators which diagonalizes the Hamiltonian as
described in Ref.\ \onlinecite{lieb61}. The surface magnetization finally
takes a simple form involving sums of products of the
couplings\cite{peschel84} 
\begin{equation}
m_s=\left(1+\sum_{j=1}^\infty\prod_{k=1}^j\lambda_k^{-2}\right)^{-1/2},
\label{ms}
\end{equation}
which stays valid for any distribution. The surface magnetization
has been evaluated in the critical region for different sequences with 
bounded fluctuations in Ref.\ \onlinecite{turban93} and in each case it
was found to vanish with a square law singularity, in agreement with
scaling  arguments. On the other hand it was shown in Ref.\ 
\onlinecite{igloi93} that any type of  marginal modulation results in a
non-universal critical behavior, with  a surface magnetization exponent
which is a continuous function of the modulation strength.

In the present paper, we continue and extend our studies of the surface 
critical behavior of aperiodic quantum Ising models. To evaluate the 
infinite sum for $m_s$ in Eq.~(\ref{ms}) we have developed a method 
leading to functional equations for the surface magnetization which 
do not contain explicitly the form of the aperiodicity. Iterating 
these relations, a closed form expression for the surface magnetization 
is obtained, from which the critical exponent as well as corrections  to
scaling can be determined exactly. To illustrate the method, 
different aperiodic sequences with irrelevant, marginal, as well as 
relevant modulations, are evaluated. For the Rudin-Shapiro sequence, which
represents a relevant perturbation, a combination of numerical results
and scaling arguments has been used. The surface magnetization  is found
to display a first order transition for some range of the  coupling
ratio, whereas the bulk transition is expected to be continuous.

The setup of the paper is the following. In Sec.~\ref{sec:aper}, the
properties of aperiodic sequences generated through substitutions are
summarized and a relevance-irrelevance criterion is deduced from scaling
considerations. Then we study successively the Thue-Morse sequence
(Sec.~\ref{sec:TM}), the period-doubling sequence
(Sec.~\ref{sec:PD}), and the Rudin-Shapiro sequence
(Sec.~\ref{sec:RS}). The results are discussed in the final section.

\section{Aperiodic sequences and scaling considerations}
\label{sec:aper}
Aperiodic sequences can be generated through iterated substitutions on
the letters $A$, $B$, \dots\  such that $A\!\rightarrow\!{\cal S}(A)$,
$B\!\rightarrow \!{\cal S}(B)$, \dots. The properties of a given sequence
are then obtained from its substitution matrix $\underline{M}$ whose
columns contain the numbers of letters $A$, $B$, \dots\  in ${\cal S}(A)$,
${\cal S}(B)$, \dots, respectively:
\begin{equation}
\underline{M}=\left(\matrix{n_A^{{\cal S}(A)}&n_A^{{\cal S}(B)}&\cdots\cr
                      n_B^{{\cal S}(A)}&n_B^{{\cal S}(B)}&\cdots\cr
                      \vdots&\vdots\cr}\right).\label{matrix}
\end{equation}
It follows that $\underline{M}^n$ has matrix elements giving the
corresponding numbers in the sequences constructed on $A$, $B$, \dots\ 
after $n$ substitutions. For example, when one starts with $A$, i.e., with
the substitutions
$
A\rightarrow{\cal S}(A)\rightarrow{\cal S}\left({\cal
S}(A)\right)\ {\rm \dots},
$
the number of $A$, $L_n^A$, in the sequence after $n$ steps is
$\left(\underline{M}^n\right)_{11}$ whereas the length of the sequence is 
$L_n\!=\!\sum_i\left(\underline{M}^n\right)_{i1}$.
Making use of the right eigenvectors ${\rm V}\!\!_\alpha$ of the
substitution matrix 
\begin{equation}
\underline{M}{\rm V}\!\!_\alpha=\Lambda_\alpha
{\rm V}\!\!_\alpha,\label{valpro}
\end{equation} 
$L_n$ and $L_n^A$ are found to be asymptotically
proportional to $\Lambda_1^n$, where $\Lambda_1$ is the largest eigenvalue
of the substitution matrix, whereas the asymptotic density of $A$ letters
involves the components of the corresponding eigenvectors
\begin{equation}
\rho_\infty^A={V_1(1)\over\sum_iV_1(i)}.\label{rhoa}
\end{equation}
Fluctuations in the numbers of letters $L_n^A$, $L_n^B$, \dots, are
connected to the next-to-leading eigenvalue $\Lambda_2$.

In the case of the aperiodic quantum Ising chain considered above, with a
two-letter sequence ($\lambda_k\!=\!\lambda_A,\lambda_B$) and an averaged
coupling $\overline\lambda$, one may write 
\begin{equation}
\lambda_A=\overline\lambda+\rho_\infty^B\delta,\qquad
\lambda_B=\overline\lambda-\rho_\infty^A\delta,\label{AB}
\end{equation}
where $\delta=\lambda_A-\lambda_B$.
Then, at a length scale $L_n\sim\Lambda_1^n$, the sum of the
deviations from the averaged coupling is
\begin{equation}
\Delta(L_n)=\sum_{k=1}^{L_n}\left(\lambda_k-\overline{\lambda}\right)
\simeq\delta\Lambda_2^n\simeq\delta L_n^\omega\label{Del}
\end{equation}
and involves the Òwandering exponent"\cite{dumont90,omega}
\begin{equation}
\omega={\ln\vert\Lambda_2\vert\over\ln\Lambda_1}.\label{wander}
\end{equation}
When the number of letters is greater than 2, one may have two
complex conjugate next-to-leading eigenvalues and the power law in
Eq.~(\ref{Del}) is modified by a factor which is periodic in $n\simeq\ln
L_n/\ln\Lambda_1$.

The aperiodicity introduces a thermal perturbation above
$\overline\lambda$ which, at a length scale $L$, is given on the average
by
\begin{equation}
\overline{\delta\lambda}(L)={\Delta(L)\over L}=\delta L^{\omega-1}.
\label{dellamb}
\end{equation}
Under a change of scale by a factor $b=L/L'$ this transforms according to
\begin{equation}
\overline{\delta\lambda'}(L')=\delta'\left(L\over
b\right)^{\omega-1}\!\! =b^{1/\nu}\delta L^{\omega-1},\label{scaling}
\end{equation}
where $\nu$ is the bulk correlation length exponent. It follows that the
perturbation amplitude $\delta$ scales like\cite{luck93a,igloi93}
\begin{equation}
\delta'=b^{\Phi/\nu}\delta,\label{delta}
\end{equation}
with a crossover exponent
\begin{equation}
\Phi=1+\nu(\omega-1).\label{crossover}
\end{equation}
As a consequence, for the two-dimensional Ising model with $\nu\!=\!1$, 
the aperiodic modulation is a relevant (irrelevant) perturbation when 
$\omega\!>\!0$ ($\omega\!<\!0$) and it becomes marginal when
$\omega\!=\!0$. In the latter case a nonuniversal behavior is expected.

For a relevant aperiodic modulation, the form of the singular
thermodynamical quantities can be deduced from scaling considerations.
Let us consider the surface magnetization as a function of the thermal
scaling field $t=1-(\lambda_c/\lambda)^2$ and the modulation amplitude
$\delta$. In a scale transformation it behaves as
\begin{equation}
m_s(t,\delta)=b^{-\beta_s/\nu}m_s\left(b^{1/\nu}t,b^{\Phi/\nu}
\delta\right),\label{scal1ms}
\end{equation}
where $\beta_s$ is the surface magnetization exponent. Taking the scale
factor $b$ equal to the bulk correlation length $b=\xi=t^{-\nu}$, one
obtains $m_s$ in the form
\begin{equation}
m_s(t,\delta)=t^{\beta_s}F\left({l_a\over\xi}\right),
\label{scal2ms}
\end{equation}
where the scaling function $F(x)$ involves a new characteristic length,
\begin{equation}
l_a=\vert\delta\vert^{-\nu/\Phi},\label{length}
\end{equation}
which is introduced by the aperiodicity and remains finite at the bulk
critical point.

In the following, an aperiodic sequence will be written as a succession
of digits $f_k\!=\!0$, $1$, which are images $\Im (A,B)$ of the letters
considered  before. Then, with $\lambda_1\!=\!\lambda r$,
$\lambda_0\!=\!\lambda$, one obtains $\lambda_k\!=\!\lambda r^{f_k}$ for
the $k$th coupling and the surface magnetization in Eq.~(\ref{ms}) can
be rewritten as 
\begin{equation}
m_s=\left[S(\lambda,r)\right]^{-1/2},\qquad
S(\lambda,r)=\sum_{j=0}^\infty\lambda^{-2j}r^{-2n_j},\label{sum}
\end{equation}
where
\begin{equation}
n_j=\sum_{k=1}^jf_k,\qquad n_0=0.\label{sumn}
\end{equation}
The critical coupling, $\lambda\!=\!\lambda_c$, is such
that\cite{pfeuty79}
$\lim_{L\to\infty}\prod_{k=1}^L\lambda_{kc}\!=\!\lim_{L\to\infty}
\lambda_c^Lr^{n_L}\!=\!1$, so that  \begin{equation}
\lambda_c=r^{-\rho_{\infty}},\qquad\rho_\infty=\lim_{L\to\infty}
{n_L\over L}.\label{criti}
\end{equation}

\section{Thue-Morse sequence}
\label{sec:TM}
As a first example, we consider the binary Thue-Morse
sequence,\cite{dekking83} which may  be deduced from the binary
representation of non-negative integers,   $0$, $1$, $10$, $11$,
$\dots$, by counting the sum of their digits modulo $2$, 
\begin{equation}
0\  1\  1\  0\  1\  0\  0\  1\  \dots\ .\label{tms}
\end{equation}
This sequence is also generated through the substitution
\begin{equation}
\left\{{A\rightarrow AB\atop B\rightarrow BA}\right.\quad {\rm
or}\quad\left\{{0\rightarrow 0\  1\atop 1\rightarrow 1\ 
0}\right.,\label{subtm} 
\end{equation}
with $\Im (A,B)=(0,1)$, so that one obtains, successively,
\begin{eqnarray}
&&0\nonumber\\
&&0\  1\nonumber\\
&&0\  1\  1\  0\nonumber\\
&&0\  1\  1\  0\  1\  0\  0\  1\nonumber\\
&&\underline{0}\  1\ \underline{1}\  0\ \underline{1}\  0\ 
 \underline{0}\  1\ \underline{1}\  0\ \underline{0}\  1\ 
 \underline{0}\  1\ \underline{1}\  0\nonumber\\
&&\dots\label{genertm} 
\end{eqnarray}
The corresponding substitution matrix (\ref{matrix}) has eigenvalues
$\Lambda_1=2$, $\Lambda_2=0$. The asymptotic density,
$\rho_\infty\!=\!1/2$, is deduced from Eq.~(\ref{rhoa}) and, according to
Eq.~(\ref{criti}),    
\begin{equation}
\lambda_c=r^{-1/2}.\label{critm}
\end{equation}
The form of the substitution in Eq.~(\ref{subtm}) immediately leads to
the recursion relations 
\begin{equation}
f_{2p}=1-f_p,\qquad f_{2p+1}=f_{p+1}.\label{ftm}
\end{equation}
{}From the second one, the sequence of odd terms, underlined in
(\ref{genertm}), reproduces the whole sequence. Splitting the sum giving
$n_j$ into even and odd parts and using (\ref{ftm}), one obtains
\begin{equation}
n_{2p}=p,\qquad n_{2p+1}=p+f_{p+1},\qquad p=0,1,2,\dots,\label{ntm} 
\end{equation}
so that a chain with length $L=2p$ has a density equal to the asymptotic
one, $\rho_\infty=1/2$, which explains the vanishing second
eigenvalue giving $\omega=-\infty$, i.e., no wandering in this case.  

\begin{figure}
\epsfxsize=8.6cm
\begin{center}
\mbox{\epsfbox{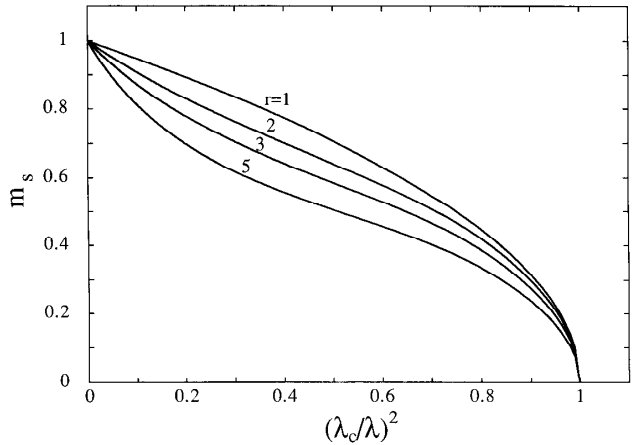}}
\end{center}
\caption{Thue-Morse surface magnetization as a function of
$1\!-\!t\!=\!(\lambda_c/\lambda)^2$ for different values of the coupling
ratio $r=\lambda_1/\lambda_0$. The critical exponent keeps its value for
the homogeneous Ising system $\beta_s=1/2$ whereas the critical amplitude
is $r$-dependent.}\label{thue-morse}  
\end{figure}

The sum $S(\lambda,r)$ involved in the surface magnetization (\ref{sum})
can be rewritten as 
\begin{eqnarray}
S(\lambda,r)&&=\sum_{p=0}^\infty\lambda^{-2(2p)}r^{-2n_{2p}}+
\sum_{p=0}^\infty\lambda^{-2(2p+1)}r^{-2n_{2p+1}}\nonumber\\
&&=\sum_{p=0}^\infty\left(\lambda^2r\right)^{-2p}+
\sum_{p=0}^\infty\lambda^{-2(2p+1)}r^{-2p-2f_{p+1}}.\label{sumtm}
\end{eqnarray}
Using the identity $a^{f_k}=1+(a-1)f_k$ ($f_k=0,1$) and the value of the
critical coupling in Eq.~(\ref{critm}), one obtains
\begin{eqnarray}
S(\lambda,r)&&={1+r\left(\lambda_c/\lambda\right)^2\over 
1-\left(\lambda_c/\lambda\right)^4}+(r^{-1}-r)
\left(\lambda/\lambda_c\right)^2\Sigma\left[
\left(\lambda_c/\lambda\right)^4\right],\nonumber\\
\Sigma(x)&&=\sum_{k=1}^\infty f_kx^k.\label{sigmatm}
\end{eqnarray}
The Thue-Morse series, $\Sigma(x)$, satisfies a recursion relation which 
follows from Eq.~(\ref{ftm}),
\begin{equation}
\Sigma(x)={x^2\over 1-x^2}+(x^{-1}-1) \Sigma(x^2),\label{sigma1}
\end{equation}
so that, iterating this relation,
\begin{equation}
\Sigma(x)=x\sum_{k=0}^\infty{x^{2^k}\prod_{p=0}^k\left(1-x^{2^p}\right)
\over\left(1-x^{2^k}\right)\left(1-x^{2^{k+1}}\right)},\label{sigma2}
\end{equation}
which is convergent for $x<1$. Equations
(\ref{sum}), (\ref{sigmatm}), and (\ref{sigma2}) give the surface
magnetization shown in Fig.~\ref{thue-morse} for any value of
$\lambda\geq\lambda_c$ .

Near the critical point, $\lambda\rightarrow\lambda_{c+}$, the leading
contribution to $S(\lambda,r)$ comes from the first two terms in
Eq.~(\ref{sigmatm}). The next ones in
$\Sigma\left[\left(\lambda_c/\lambda\right)^4\right]$ give the
corrections to scaling. Collecting these results, the surface
magnetization finally takes the form 
\begin{equation}
m_s={2t^{1/2}\over\lambda_c+\lambda_c^{-1}}\left[1+{1\over 4}\left(
{1-\lambda_c^2\over 1+\lambda_c^2}\right)^2t+O(t^2)\right],\label{mstm}
\end{equation}
where $t$ is the deviation from the critical point defined in
Sec.~\ref{sec:aper}. The surface magnetization exponent takes the
value $\beta_s=1/2$ for homogeneous Ising systems in two dimensions as
expected from scaling since $\omega<0$ here and the aperiodic modulation
of the couplings is an irrelevant perturbation. The critical amplitude
depends on $r$ through $\lambda_c$ in agreement with previous
results.\cite{turban93} 

\section{Period-doubling sequence}
\label{sec:PD}
\begin{figure}
\epsfxsize=8.6cm
\begin{center}
\mbox{\epsfbox{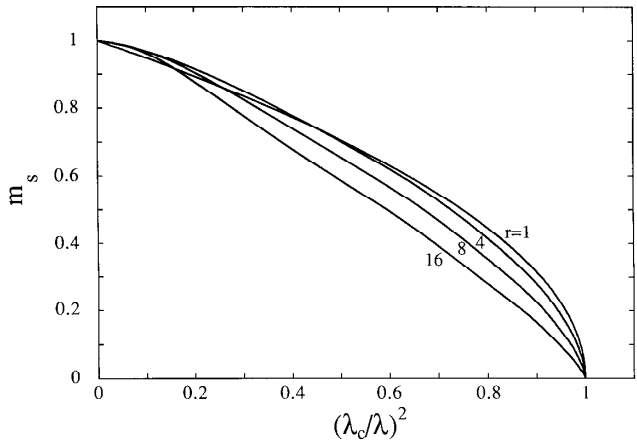}}
\end{center}
\caption{Period-doubling surface magnetization for different
values of $r$. The exponent $\beta_s$ varies
continuously with the modulation amplitude.}\label{period-doubling}
\end{figure}

We now turn to the period-doubling sequence which
follows from the substitution\cite{dekking83,luck93a}
\begin{equation}
\left\{{1\rightarrow 1\  0\atop 0\rightarrow 1\  1}\right.,\label{subpd}
\end{equation}
leading to
\begin{eqnarray}
&&1\nonumber\\
&&1\  0\nonumber\\
&&1\  0\  1\  1\nonumber\\
&&1\  0\  1\  1\  1\  0\  1\  0\nonumber\\
&&1\ \underline{0}\  1\ \underline{1}\  1\ \underline{0}\  1\ 
 \underline{0}\  1\ \underline{0}\  1\ \underline{1}\  1\ 
 \underline{0}\  1\ \underline{1}\nonumber\\
&&\dots\label{generpd}
\end{eqnarray} 
The eigenvalues of the substitution matrix are then $\Lambda_1\!=\!2$,
$\Lambda_2\!=\!-1$ and the asymptotic density $\rho_\infty\!=\!2/3$ gives
the critical coupling 
\begin{equation}
\lambda_c=r^{-2/3}.\label{cripd}
\end{equation}
The form of the substitution is such that
$f_{2p}=1-f_p$, $f_{2p+1}=1$, from which one deduces
\begin{equation}
n_{2p}=2p-n_p,\quad n_{2p+1}=2p+1-n_p,\quad p=0,1,2,\dots\  .\label{npd}
\end{equation}
Splitting the sum in $S(\lambda,r)$ into even and odd parts as in
Eq.~(\ref{sumtm}) and using (\ref{npd}), one obtains the following
recursion relation: 
\begin{equation}
S(\lambda,r)=\left(1+{1\over\lambda^2r^2}\right)S(\lambda^2r^2,r^{-1}).
\label{recurpd}
\end{equation}
After $l$ iterations the arguments in $S(\lambda,r)$ are changed into
\begin{eqnarray}
\lambda_l&&=\left\{\matrix{
\lambda^{2^l}r^{{2\over 3}(2^l+1)},&&l&{\rm odd},\cr
\lambda^{2^l}r^{{2\over 3}(2^l-1)},&&l&{\rm even},\cr}
\right.,\nonumber\\
r_l&&=r^{(-1)^l},\qquad l=0,1,2,\dots,\label{coupling}
\end{eqnarray}
and the series can be written as an infinite product
\begin{equation}
S(\lambda,r)=\prod_{k=1}^\infty\left[1+\lambda_c
{\left(\lambda_c\over\lambda\right)}^{2^{2k-1}}\right]\left[1+\lambda_c^{-1}
{\left(\lambda_c\over\lambda\right)}^{2^{2k}}\right],\label{prodpd}
\end{equation}
which, together with (\ref{sum}), gives the surface magnetization shown
in Fig.~\ref{period-doubling}.

In order to analyze the critical behavior of the surface magnetization,
one may use a scaling method introduced by one of us.\cite{igloi86} Let
$S(u)$ denote the series expansion of $S(\lambda,r)$ in powers of
$u=(\lambda_c/\lambda)^2$. According to Eq.~(\ref{sum}), $S(u)$ should display a
power law singularity near the
critical point:  
\begin{equation}
S(u)\sim (1-u)^{-2\beta_s}.\label{scrit}
\end{equation}
It may be shown that in this case, the truncated series $S_L(u)$ given by
the first $L$ terms in $S(u)$ behaves, for large values of $L$, as
$L^{2\beta_s}$ at the critical point, $u=1$. Now it may be
verified that the first $l$ terms in the infinite product (\ref{prodpd})
just contain the $L=2^{2l}$ first terms of the series expansion. As a
consequence, one obtains 
\begin{equation}
S_{L=2^{2l}}(1)=\left[(1+\lambda_c)(1+\lambda_c^{-1})\right]^l\sim
\left(2^{2l}\right)^{2\beta_s},\label{scals}
\end{equation}
from which the surface magnetization exponent 
\begin{equation}
\beta_s={\ln\left[(1+\lambda_c)(1+\lambda_c^{-1})\right]\over 4\ln
2}
\label{betapd}
\end{equation}
follows. 

The same result can be obtained by considering the recursion relation
(\ref{recurpd}) at the next step:
\begin{equation}
S(\lambda,r)=\left(1+{1\over\lambda^2r^2}\right)\left(1+{1\over
\lambda^4r^2}\right)S(\lambda^4r^2,r).\label{recur2}
\end{equation}
At this stage the new coupling on the right-hand side is
$\lambda'\!=\!\lambda^4r^2$ whereas $r$ and, as a consequence, $\lambda_c$
remain unchanged. When $\lambda\to\lambda_c$, $S(\lambda,r)$ behaves as in
Eq.~(\ref{scrit}) with an amplitude $A(r)$ and, in the same way,
\begin{eqnarray}
S(\lambda^4r^2,r)&\simeq& A(r)\left[1-\left({\lambda_c\over\lambda'}
\right)^2 \right]^{-2\beta_s}\nonumber\\
&\simeq& A(r)\left[1-\left({\lambda_c\over\lambda}\right)^8
\right]^{-2\beta_s},\label{scal2} 
\end{eqnarray}
where the value of $\lambda_c$ given in Eq.~(\ref{cripd}) was used.
Introducing these results into (\ref{recur2}) an equation for $\beta_s$
is obtained which leads to (\ref{betapd}). 

\begin{figure}
\epsfxsize=8.6cm
\begin{center}
\mbox{\epsfbox{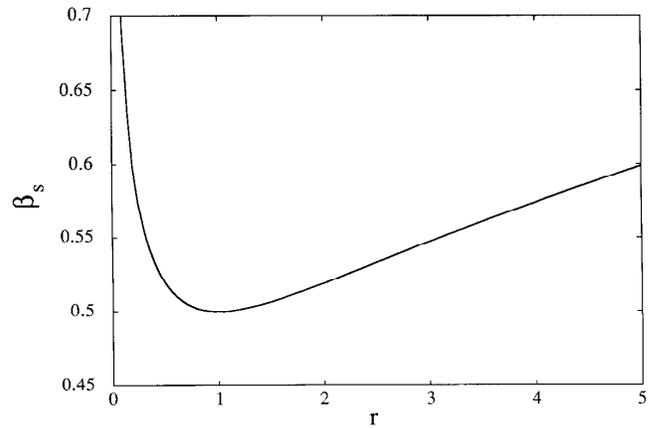}}
\end{center}
\caption{Marginal variation of surface exponent $\beta_s$
with the coupling ratio $r$ for the period-doubling sequence.
In the aperiodic system the singularity is always weaker than in the
homogeneous one and the surface transition is of second order for
any value of $r$.} 
\label{betas-marginal}
\end{figure}

The surface magnetization exponent depends on the amplitude of the
aperiodicity $r$ through $\lambda_c$ as expected for this sequence 
since $\omega$ vanishes and the perturbation is marginal. The variation
is shown in Fig.~\ref{betas-marginal}.

Let $m_s(L)$ be the matrix element in Eq.~(\ref{mat-el}) for a chain with
length $L$. Although the spontaneous magnetization vanishes on a finite
chain, $m_s(L)$ remains nonvanishing and scales with $L$ as
$L^{-x_s}$ at the critical point\cite{hamer82,takano85} where 
$x_s=\beta_s/\nu$ is the scaling dimension of the surface magnetization.
To the leading order in $L$, $m_s(L)$ is still given by Eq.~(\ref{ms}) with a
sum over $j=1,L$ so that the truncated series $S_L(1)$ considered above also
behaves as $L^{2x_s}$. It follows that $\beta_s\!=\!x_s$ and $\nu\!=\!1$: The
marginal aperiodic modulation does not change the bulk correlation length
exponent.

\section{Rudin-Shapiro sequence}
\label{sec:RS}
As a final example, we consider the Rudin-Shapiro
sequence,\cite{dekking83,luck93a} which is generated using a four-letter
substitution 
\begin{equation}
\left\{\matrix{A&\rightarrow AB\cr
               B&\rightarrow AC\cr
               C&\rightarrow DB\cr
               D&\rightarrow DC\cr}\right..\label{subrs}
\end{equation}
One may take either single-digit images of the four letters in
(\ref{subrs}) with\cite{luck93a} $\Im(A,B,C,D)\!=\!(0,0,1,1)$ or, more
naturally, two-digit images with $\Im(A,B,C,D)\!=\!(00,01,10,11)$ so
that
\begin{equation}
\left\{\matrix{&00\rightarrow 0001\cr
               &01\rightarrow 0010\cr
               &10\rightarrow 1101\cr
               &11\rightarrow 1110\cr}\right..\label{sub2rs}
\end{equation}
The two processes lead to the same sequence of $1$ and $0$ since the
two-digit images are related to the single-digit ones through a
substitution on the letters. Starting on $A$, one obtains
\begin{eqnarray}
&&0\  0\nonumber\\
&&0\  0\  0\  1\nonumber\\
&&0\  0\  0\  1\  0\  0\  1\  0\nonumber\\
&&\underline{0}\  0\ \underline{0}\  1\ \underline{0}\  0\ 
 \underline{1}\  0\ \underline{0}\  0\ \underline{0}\  1\ 
 \underline{1}\  1\ \underline{0}\  1\nonumber\\
&&\dots\label{generrs} 
\end{eqnarray}
The substitution matrix has leading eigenvalues $\Lambda_1\!=\!~2$,
$\Lambda_2\!=\!\pm\sqrt{2}$ so that $\omega\!=\!1/2$ and the asymptotic
density $\rho_\infty=1/2$, obtained by taking a weighted average on the
densities for the different letters, gives the critical coupling
\begin{equation}
\lambda_c=r^{-1/2}.\label{crirs}
\end{equation} 
{}From the form of the substitution in Eq.~(\ref{sub2rs}), one easily
deduces the following relations: 
\begin{equation} 
f_{4p}=1-f_{2p},\quad f_{4p+1}=f_{4p+2}=f_{2p+1},\quad
f_{4p+3}=f_{2p+2}.\label{frs} 
\end{equation}
Furthermore, one may notice that odd terms, underlined in
(\ref{generrs}), reproduce the sequence itself whereas even terms
either reproduce the sequence or its complement to one, so that:
\begin{equation}
f_{2p+1}=f_{p+1},\qquad
f_{2p+2}=\left\{\matrix{1-f_{p+1},&&p&{\rm odd},\cr      
                          f_{p+1},&&p&{\rm even}.\cr}\right.
\label{f2rs} 
\end{equation}

\begin{figure}
\epsfxsize=8.6cm
\begin{center}
\mbox{\epsfbox{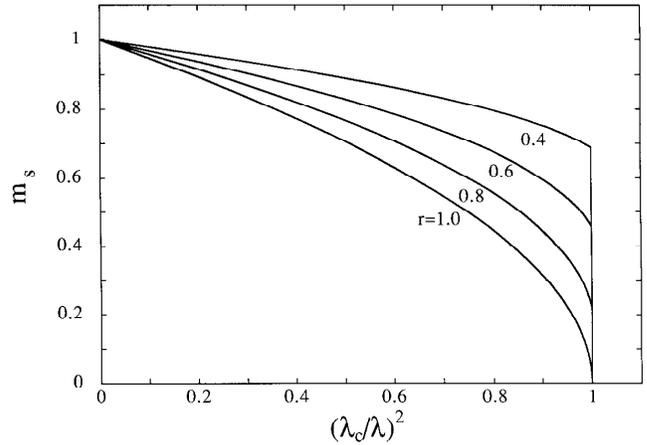}}
\end{center}
\caption{Surface magnetization for the Rudin-Shapiro sequence with a
coupling ratio $r\leq 1$. The enhancement of the coupling strength  near
to the surface leads to a first-order surface transition when $r<1$.} 
\label{first-order}
\end{figure}

\begin{figure}
\epsfxsize=8.6cm
\begin{center}
\mbox{\epsfbox{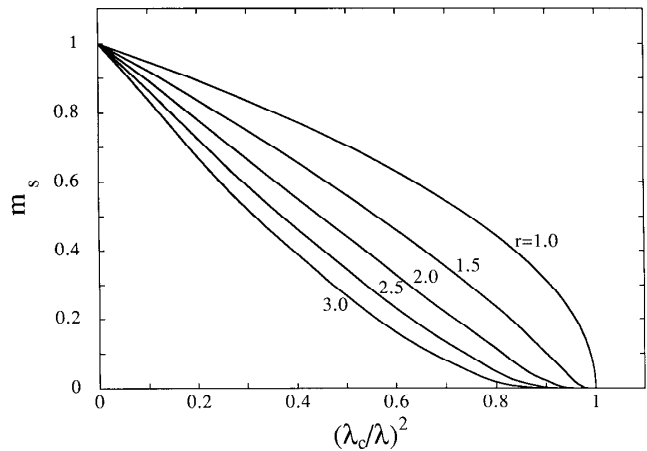}}
\end{center}
\caption{Surface magnetization for the Rudin-Shapiro sequence with
$r\geq 1$. The surface magnetization vanishes with an
essential singularity at the bulk critical point when $r>1$.}
\label{essential}
\end{figure}

Using these properties one may write a recursion relation similar 
to~(\ref{recurpd}), but here in matrix form, between the odd and
even parts of $S(\lambda,r)$. Although the matrix considerably
simplifies at the bulk critical point, some of its elements still depend
on the iteration index $l$. As a consequence, we were unable to deduce the
critical behaviour from this recursion relation like in
Sec.~\ref{sec:PD}.

Numerical results for the surface magnetization are shown in
Figs.~\ref{first-order} and~\ref{essential} for $r<1$ and $r>1$,
respectively. In the first case, one obtains a first order surface
transition, which is linked to a local enhancement of the couplings
relative to their averaged value. 
On the other side, when $r>1$, the couplings near the surface are locally
weaker than the average and the surface magnetization vanishes at the
bulk critical point with a singularity which is seemingly weaker than any
power law. In the bulk, the situation is different since the
deviations from the average have both signs with the same probability.

That the first order transition is indeed a surface effect can also be
made plausible through the following argument: If one started the
sequence with $D$ instead of $A$, the bulk properties should not change
but one can check on Eqs.~(\ref{subrs}) and ~(\ref{sub2rs}) that it
would amount to exchange weak and strong couplings along the sequence,
and the first order surface transition would then occur for $r>1$.

\begin{figure}
\epsfxsize=8.6cm
\begin{center}
\mbox{\epsfbox{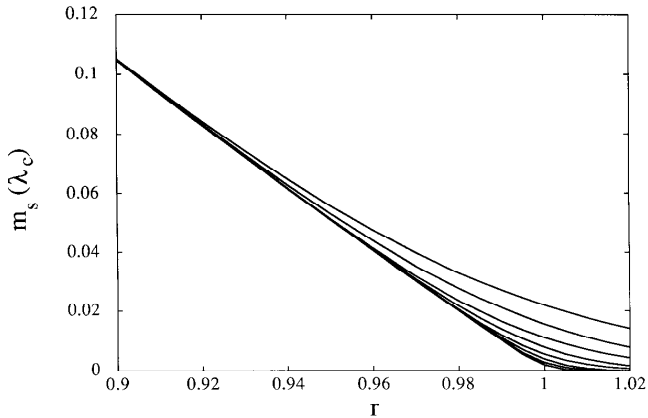}}
\end{center}
\caption{Surface magnetization discontinuity at the bulk critical
point as a function of the coupling ratio for the Rudin-Shapiro sequence
with increasing sizes, $L=2^{11}$ to $2^{18}$, from top to bottom. The
variation is asymptotically linear~in~$r$.} \label{discontinuity} 
\end{figure}

The discontinuity of the critical surface magnetization shown in
Fig.~\ref{discontinuity} appears to vary linearly with the coupling
ratio $r<1$. This behavior will be explained in
Sec.~\ref{sec:discussion} using scaling arguments.  

\section{Discussion}
\label{sec:discussion}
The QIM on aperiodic lattices studied in this paper shows a rich
critical behavior. Depending on the value of the wandering exponent
$\omega$, scaling arguments show that the perturbation introduced by the
aperiodic modulation can be either irrelevant, marginal or relevant.
This has been verified on three specific sequences for which the
different typical behaviors are observed.

For the period-doubling sequence, displaying a marginal modulation,
the surface critical exponent $\beta_s$ and the corresponding
scaling dimension $x_s=\beta_s/\nu$ are nonuniversal and show the same
continuous variation with the strength of the modulation. As a
consequence $\nu$ itself stays constant, keeping its value in the
homogeneous system, $\nu=1$. Thus the marginality condition
$1/\nu=1-\omega$ is fulfilled along the critical line, as required. On
the other hand one knows from numerical studies\cite{luck93a} that the
bulk specific heat exponent $\alpha$ does vary with $\delta$ so that the
hyperscaling relation $d\nu=2-\alpha$ is violated for this marginal
sequence. Such a violation of hyperscaling is likely to occur for other
marginal sequences too. It may be due to the presence of a dangerous
irrelevant variable in the problem.\cite{fisher74}

\begin{figure}
\epsfxsize=8.6cm
\begin{center}
\mbox{\epsfbox{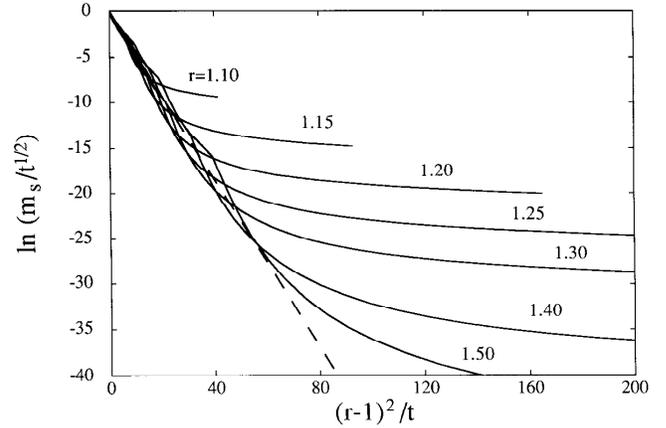}}
\end{center}
\caption{Logarithm of the reduced surface magnetization as a function
of $x^{-1}$ for the Rudin-Shapiro sequence with $r>1$ and $L=2^{16}$. 
The linear variation (dashed line) is in agreement with the stretched
exponential form in Eq.~(\protect\ref{stretched}). Deviations from
scaling are mainly due to finite-size effects.}
\label{scaling-es}
\end{figure}

With the Rudin-Shapiro sequence, the relevant modulation was found to
lead to different surface critical behaviors, depending on whether the
coulings near to the surface are greater or smaller than their average.
In the first case, the surface stays ordered at the bulk critical point.
The variation with $r$ of surface magnetization discontinuity
can be deduced from the scaling form in Eq.~(\ref{scal2ms}). For
$m_s(0,\delta)$ to remain finite at the bulk critical point, the
$t$ dependence on the right-hand side has to cancel. The scaling function then
behaves as a power of its argument which follows from the cancellation,
leading to 
\begin{equation}
m_s(0,\delta)\sim
l_a^{-x_s}\sim\vert\delta\vert^{\beta_s/\Phi}.
\label{scaldisc} \end{equation} 
In the present case the exponent is equal to $1$ and the discontinuity
varies as $\vert\delta\vert\!=\!1-r$ in agreement with the numerical
results in~Fig.~\ref{discontinuity}.

The same behavior is obained in the the Hilhorst--van Leeuwen (HvL)
model\cite{hilhorst81} in the case of an
inhomogeneity decaying as $AL^{-y}$ where $L$ is the distance from the
free surface. In this case the crossover exponent is $1-\nu y$ so that,
according to~(\ref{crossover}), $y$ corresponds to $1-\omega$ which, in
the present problem, is the decay exponent for the average deviation of
the couplings given in Eq.~(\ref{dellamb}). The jump of the surface
magnetization in the Ising HvL model has been obtained
analytically\cite{peschel84} for a relevant perturbation with $A>0$ and
is in agreement with~(\ref{scaldisc}).

When $r>1$, the local couplings at the surface are in average weaker than
the critical one and, according to the numerical results shown in
Fig.~\ref{essential}, the surface order vanishes in an anomalous way.
In order to obtain the form of the scaling function $F(x)$ in
Eq.~(\ref{scal2ms}), one may use the abovementioned analogy with the HvL
model as done in~Ref.~\onlinecite{igloi93}. An argument of
Burkhardt,\cite{burkhardt82} going beyond scaling theory, leads
to a stretched exponential behavior for the critical correlation
function, from which the leading singularity of the susceptibility can
be deduced,\cite{igloi93} 
\begin{equation}
\chi\sim\exp\left[-C_sx^{\Phi/(\Phi-1)}\right].\label{susc}
\end{equation}
In this expression $x=l_a/\xi$ is the scaling variable defined
in~Eq.~(\ref{scal2ms}). The same form is likely to occur in the temperature
dependence of other singular quantities and one expects the reduced surface
magnetization to behave as \begin{equation}
{m_s\over t^{\beta_s}}\sim\exp\left[-C_mx^{1+1/\nu(\omega-1)}\right],
\label{stretched}
\end{equation}
where $C_m$ is some constant. The scaling function may also involve
some unknown power of $x$ in front of the exponential. This
expression, when properly translated, is in agreement with the analytical
result obtained by Peschel for the surface
inhomogeneity.\cite{peschel84} For the aperiodic Ising model with
Rudin-Shapiro modulation, $\omega=1/2$ and $\nu=1$, so that the argument
of the exponential in Eq.~(\ref{stretched}) is $x^{-1}\!=\!(r-1)^2/t$.
The numerical results shown in Fig.~\ref{scaling-es}, although perturbed
by finite-size effects, confirm the proposed scaling form.

All these systems containing a characteristic length, which, like $l_a$, 
stays finite at the bulk critical point when the perturbation is
relevant, seem to display the same type of critical  behaviour. Besides
the two examples considered so far, one may also mention 
systems limited by a surface with a parabolic shape, for which the
characteristic length is fixed by the geometry.\cite{IPT93} They also
differ in some aspects: In the marginal HvL model, for instance, the
surface transition is first order when the amplitude of
the inhomogeneity is positive and strong enough, whereas it remains second
order for the period-doubling modulation studied here, whatever the
modulation amplitude.

\acknowledgments
The authors are indebted to J. M. Luck for communicating his work,
prior to publication. This work was supported by the C.N.R.S. and the
Hungarian Academy of Sciences through an exchange program. F.I. gratefully
acknowledges support by the National Hungarian Research Fund under grant No
OTKA TO12830. The Laboratoire de Physique du Solide is Unit\'e de Recherche
Associ\'ee au C.N.R.S. No 155.


\begin{references}
\bibitem{omega}We abandon the usual notation $\beta$ for this exponent
in order to avoid a confusion with the magnetization exponent.
\end{references}
\end{document}